\documentclass{article}
\usepackage{spconf,amsmath,graphicx}
\usepackage[final,markup=underlined]{changes}
\definechangesauthor[color=green]{Aditay}

\title{Adversarial Learning of Raw Speech Features for Domain Invariant Speech Recognition}
\name{Aditay Tripathi  $^{\star \ddagger}$ \qquad Aanchan Mohan $^\dagger$ \qquad Saket Anand $^\star$ \qquad Maneesh Singh $^\ddagger$}
\address{$^\star$ Indraprastha Institute of Information Technology, New Delhi, India.\\
         $^\dagger$ Synaptitude Brain Health, Vancouver, Canada.\\
         $^\ddagger$ Verisk Analytics, Jersey City, USA.
         }
\begin{document}
	\ninept
	\maketitle
	\begin{abstract}
		Recent advances in neural network based acoustic modelling have shown significant improvements in automatic speech recognition (ASR) performance. In order for acoustic models to be able to handle large acoustic variability, large amounts of labeled data is necessary, which are often expensive to obtain. This paper explores the application of adversarial training to learn features from \textit{raw speech} that are invariant to acoustic variability. This acoustic variability is referred to as a domain shift in this paper. The experimental study presented in this paper leverages the architecture of Domain Adversarial Neural Networks (DANNs)~\cite{dann} which uses data from two different domains. The DANN is a Y-shaped network that consists of a multi-layer CNN feature extractor module that is common to a label (senone) classifier and a so-called domain classifier. The utility of DANNs is evaluated on multiple datasets with domain shifts caused due to differences in gender and speaker accents. Promising empirical results indicate the strength of adversarial training for unsupervised domain adaptation in ASR, thereby emphasizing the ability of DANNs to learn domain invariant features from raw speech. 
	\end{abstract}
    \vspace{-0.15in}
	\section{Introduction}
	\label{sec:intro}
    \vspace{-0.08in}
Training neural network based acoustic models for Automatic Speech Recognition (ASR) gets challenging when limited amounts of supervised training data is available. It is expensive to obtain labeled speech data that contains sufficient variations of the  different sources of acoustic variability such as speaker accent, speaker gender, speaking style, different types of background noise or the type of recording device~\cite{benzeghiba2007automatic}. To mitigate the effects of acoustic variability that is inherent in the speech signal, domain adaptation techniques are often used in acoustic modelling. This paper investigates the use of Domain Adversarial Neural Networks (DANNs)~\cite{dann} for domain adaptation of the acoustic model from raw speech directly instead of relying on traditional log-mel features. Although MFCC or log-mel features are predominantly used in acoustic modeling, there has been significant recent interest to learn acoustic models using raw speech~\cite{palaz2015analysis,palaz2015convolutional,conf/interspeech/PalazCM13,ghahremani2016acoustic}. For example, Sainath et.al.~\cite{sainath2015learning} showed that recognition performance on raw speech matches the performance on log-mel filters. 
	
	Several techniques have been proposed to mitigate the effects of acoustic variability in the speech data at test time. 
	Feature space maximum
	likelihood linear regression (fMLLR)~\cite{gales2008application}, Maximum Likelihood Linear Regression~\cite{gales2008application}, MAP~\cite{oh2009mllr}, Vocal Tract Length Normalization (VLTN)~\cite{zhan1997vocal} are well-known speaker adaptation techniques used in generative acoustic models. i-Vectors~\cite{gupta2014vector}, LHUC~\cite{swietojanski2014learning} and KL-divergence regularized DNN acoustic models\cite{yu2013kl} are the popular adaptation techniques used for discriminative acoustic models. All of these techniques require labeled data from the target domain to perform adaptation. 
	
	The success of adversarial training using DANNs for unsupervised domain adaptation in computer vision~\cite{dann} suggests promising extensions to ASR applications as well. In a DANN, domain adaptation is achieved by incorporating the additional task of domain classification along with label classification. Both the domain-classifier and the label (senone) classifier share a common multi-layer CNN feature extraction module. The network is trained to \emph{minimize} the cross-entropy loss of the label classifier and at the same time \emph{maximize} that of the domain classifier. DANN has been used to learn domain-invariant feature representations, thus achieving unsupervised domain adaptation for acoustic models trained on log-mel features  ~\cite{shinohara2016adversarial,serdyuk2016invariant,sun2017unsupervised}. Variational Autoencoders (VAE)~\cite{hsu2017unsupervised} have also been applied for unsupervised domain adaption. These techniques have been studied for domain adaptation for discriminative acoustic models trained on log filter-bank features. This paper presents an experimental study of unsupervised domain adaptation on discriminative acoustic models trained on raw speech by using DANNs. Unsupervised domain adaptation is used to reduce acoustic variability due to (1) speaker gender and (2) speaker accent. To study the impact of DANN unsupervised domain adaptation on acoustic variability arising due to variations in speaker gender, experimental results are presented on the TIMIT data set. Furthermore, British and American accented data from the Voxforge corpus is used to study the impact of DANN unsupervised domain adaptation on acoustic variability arising due to differing speaker accents. The experimental study in this paper shows that domain invariant features can be learned directly from raw speech with significant improvement over the baseline acoustic models trained without domain adaptation.
    
    The remainder of the paper is organized as follows. Section 2 reviews related work done in the area of domain-invariant feature learning using domain adversarial neural networks. Section 3 describes the adversarial training for unsupervised domain adaptation for ASR trained on raw speech and Section 4 presents the experimental setup, description of input features and a description of the configuration of the acoustic model. Section 5 presents the experimental results obtained followed by a discussion in Section 6 and conclusion in Section 7.    
    
    \vspace{-0.15in}
    \section{Relation to prior work}
    \vspace{-0.08in}
    \label{sec:previous work}
    Domain adaptation using adversarial training was first introduced by Ganin et.al~\cite{dann} for domain adaptation in computer vision. It has since been used for noise invariant feature learning in ASR using supervised labels and filter-bank features~\cite{shinohara2016adversarial,serdyuk2016invariant}. Sun et.al.~\cite{sun2017unsupervised} used adversarial training for unsupervised domain adaptation for robust speech recognition using filter bank features and used the WSJ and Librispeech speech corpora as data from different domains. All of these studies used filter-bank features to learn the domain invariant features for ASR. In this study, domain invariant features are learned directly from raw speech to mitigate acoustic variabilities due to speaker gender and accent which often adversely affect ASR performance ~\cite{huang2004accent,tatman2017effects}.
\vspace{-0.15in}
\section{Domain adaptation using raw speech features}
	\label{sec:format}
\begin{figure*}[!t]
		\centering	
		\includegraphics[scale=0.3]{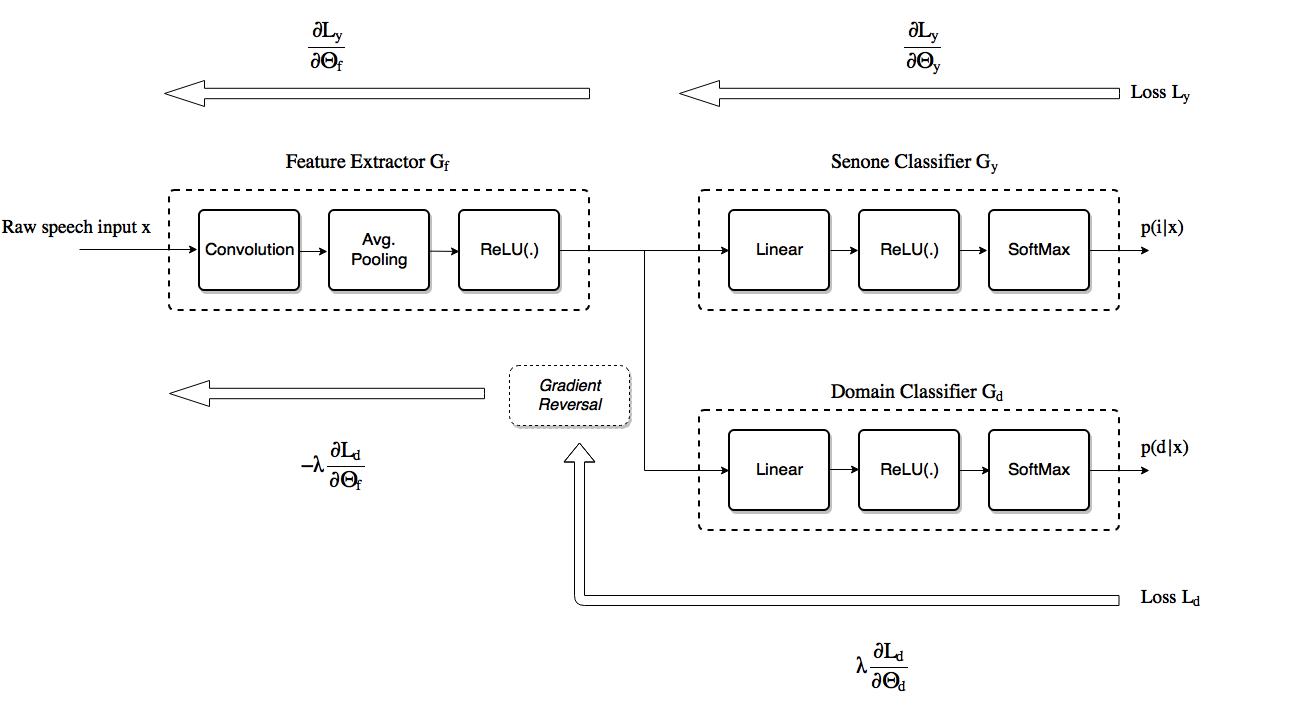}
        \vspace{-0.2in}
		\caption{ The feature extractor extracts discriminative features directly from raw speech. It consists of several stages of convolution/pooling/ReLU. Senone classifier along with the feature extractor forms a standard architecture. Domain classifier is also connected to feature extractor through a gradient reversal. Reversal of gradient force the feature extractor to learn domain invariant features.}
        \label{fig:dann2}
        \vspace{-0.2in}
	\end{figure*}

    \vspace{-0.011in}
    This section explains unsupervised domain adaptation using adversarial training on raw speech features. Consider a classification problem with the input feature vector space $X$ and $Y=\{0,1,2,...,L-1\}$ as the set of labels in the output space. Let $S(x,y)$ and $T(x,y)$ be \emph{unknown} joint distributions defined over $X \times Y$, referred to as the source  and target distributions respectively. The unsupervised domain adaptation algorithm requires input as the \emph{labeled source domain data}, sampled from $S(x,y)$ and \emph{unlabeled target domain data}, sampled from the marginal distribution $T(x)$ i.e.	
	\begin{equation}
		\{(x_i,y_i)\}_{i=0}^{n} \sim {S(x,y)};  \; \{(x_i)\}_{i=n+1}^{n+n'=N} \sim T(x), \nonumber
	\end{equation}
	where $N=n+n'$ is the total number of input samples. As opposed to the class labels, which are assumed only for the source domain data, the \emph{binary} domain labels $(d_i=\{0,1\})$ are defined as 
	\begin{equation}
	d_i=
		\begin{cases}
		0  \text{ for $x_i\sim S(x,y)$}\\
		1 \text{ for $x_i\sim T(x)$.}
		\end{cases}
	\end{equation}
and are assumed to be known for each sample.
	The neural network architecture is as shown in Fig. \ref{fig:dann2} and comprises of three modules: the feature extractor, label classifier and the domain classifier. The feature extractor is a multi-layer Convolutional Neural Network (CNN) that takes as input $x_i$, while its output is available to the label and domain classifiers, which predict the labels $y_i$ and $d_i$ respectively. At training time, the label classifier's loss is only computed over labeled samples from $S(x,y)$, whereas the domain classifier's loss is computed over both, labeled samples from $S(x,y)$ and unlabeled samples from $T(x)$. The feature extractor and the two classifier modules are described in detail in Sections \ref{sec:feature extractor} and \ref{sec:classifiers}. 
    \vspace{-0.15in}
    
	\subsection{Feature Extractor}
	\label{sec:feature extractor}
    \vspace{-0.05in}
	The feature extractor, $G_f$, shown in Fig. \ref{fig:dann2} is a multi-layer CNN and takes the raw speech input vector $x_i$ and generates a $d$-dimensional feature vector $f_i\in R^d$ i.e.,
	\begin{equation}
		f_i = G_f(x_i; \Theta_f),		
	\end{equation}
	where $\Theta_f$ are the parameters of the feature extractor i.e. weights and biases of the convolutional layers. The input vector $x_i$ can be from the source distribution $S(x,y)$ or the target distribution $T(x)$. The 1-d convolution operation in the convolutional layer in the network is defined as below
	\begin{equation}
		\label{eq:conv}
		f_{i}^{m,c,1}=\sigma(\sum_{j=m}^{m+k-1}\theta_{f}^{j-m,c,1}\cdot x_i^j),
	\end{equation}	
 Eq. (\ref{eq:conv}) gives feature vector output at index $m$ from the first layer convolution operation on input feature vector $x_i$, $\theta_{f}^{c,1}$ denotes the $k$-dimensional vector of weights and biases of the first convolutional layer and $c^{th}$ convolutional filter. The function $\sigma(\cdot)$ is a non-linear activation function like the sigmoid or ReLU.  
	\vspace{-0.15in}
	\subsection{Label and Domain Classifiers}
	\label{sec:classifiers}
	The feature vector $f_i$, which is extracted from $G_f$, is mapped to class label ${y}_i=G_y(f_i;\Theta_y)$ by the label classifier $G_y\, $ and to domain label ${d}_i=G_d(f_i;\Theta_d)$ by a domain classifier $G_d$ as shown in Fig. \ref{fig:dann2}. Both the label classifier as well as domain classifier are multi-layer feed-forward neural networks with parameters collectively denoted as $\Theta_f$ and $\Theta_d$ respectively. The unsupervised domain adaptation is achieved by training the network to minimize the cross-entropy based label classification loss on the labeled source domain data and at the same time \emph{maximize} the cross-entropy domain classification loss on the supervised source domain data and unsupervised target domain data. The classification losses is the cross-entropy costs. The total loss is given by
		\begin{equation}
	E(\Theta_f, \Theta_y, \Theta_d)=\sum_{i=1..N,d_i=0}^{}L_y(G_y(G_f(x_i;\Theta_f);\Theta_y), y_i)-\nonumber
	\end{equation}
\vspace{-0.1in}	\begin{equation}\lambda\sum_{i=1..N}L_d(G_d(G_f(x_i;\Theta_f);\Theta_d),d_i).
	\end{equation}
	The parameter $\lambda$ is a hyper-parameter that weighs the relative contribution of the two costs. To simplify the above equations, these are written in the compressed form as below 
		\begin{equation}
		E(\Theta_f,\Theta_y,\Theta_d)=\sum_{i=1..N,d_i=0}L^i_y(\Theta_f, \Theta_y)-\nonumber
		\end{equation}
        \vspace{-0.1in}
		\begin{equation}
		\lambda\sum_{i=1..N}L^i_d(\Theta_f, \Theta_d).
		\end{equation}
The label classifier tries to minimize the label classification loss $L_y^i(\Theta_f, \Theta_y)$ on the data from source distribution $S(x,y)$, therefore the parameters of both feature extractor $(\Theta_f)$ and label predictor$(\Theta_y)$ are optimized. This ensures that the features $f_i$ are discriminative enough to perform good prediction on samples from the source domain. At the same time the extracted features should be invariant to the shift in domain. In order to obtain domain invariant features, the parameters of feature extractor $\Theta_f$ are optimized to maximize the domain classification loss $L_y(\Theta_f,\Theta_d)$ while at the same time domain classifier $\Theta_d$ tries to classify the input features. In other words, the domain classifier of the trained network should not be able to correctly predict the domain labels of the features coming from the feature extractor. 
	
	
The desired parameters $\hat{\Theta_f},\hat{\Theta_y}, \hat{\Theta_d}$ give the saddle point at the training and are estimated as:
	\begin{equation}
	(\hat{\Theta _f}, \hat{\Theta_y})=arg\,\operatorname*{min}_{\Theta _f,\Theta_y}E(\Theta_f, \Theta_y,\hat{\Theta_d})
	\end{equation}
	\vspace{-0.1in}
	\begin{equation}
	\hat{\Theta_d} = arg\,\operatorname*{max}_{\Theta_d}E(\hat{\Theta_f}, \hat{\Theta_y}, \Theta_d).
	\end{equation}
	The model can be optimized by the standard stochastic gradient descent (SGD) based approaches. The parameter updates during the SGD becomes	
	\begin{equation}
	\label{eq:grl}
	\Theta_f \leftarrow \Theta_f - \mu\Bigl\{\frac{\partial L_y^i}{\partial \Theta_f}-\lambda \frac{\partial L_d^i}{\partial \Theta_f}\Bigr\}
	\end{equation}
    \vspace{-0.1in}
	\begin{equation}
	\label{eq:grl2}
	\Theta_y \leftarrow \Theta_y - \mu \frac{\partial L_Y^i}{\partial \Theta_y}
	\end{equation}
    \vspace{-0.1in}
	\begin{equation}
	\label{eq:grl3}
	\Theta_d \leftarrow \Theta_d - \mu \frac{\partial L_d^i}{\partial \Theta_d}.
	\end{equation}	
	where, $\mu$ is the learning rate. Eq. \ref{eq:grl}-\ref{eq:grl3} can be implemented in some form of SGD by using a special Gradient Reversal Layer (GRL) at the end of feature extractor and at the beginning of domain classifier as shown in fig. \ref{fig:dann2}. During the backward propagation, GRL reverses the sign of  gradients, multiply them with $\lambda$ and pass it onto the subsequent layer, while in forward propagation GRL acts as an identity transform.
	
	At the test time, domain classifier and GRL are discarded. The data samples are passed through the feature extractor and label classifier to get the predictions.      
	\vspace{-0.15in}
	\section{EXPERIMENTAL SETUP}
	\label{sec:page}
    \vspace{-0.08in}    
    This section explains the setup used for domain adaptation experiments. Sec. \ref{sec:database} describes the dataset used for the experiments, sec. 
   \ref{sec:features} details the input feature preparation and sec. \ref{sec:model} discusses the model architecture and training details. 
   \vspace{-0.15in}
	\subsection{Dataset Description}
	\label{sec:database}
    \vspace{-0.08in}	
	TIMIT~\cite{garofolo1993darpa} and Voxforge~\cite{maclean2012voxforge} datasets are used to perform domain adaptation experiments. For TIMIT speech corpus domain adaptation is performed by taking male speech as source domain and female speech corpus as target domain. For the Voxforge corpus domain adaptation is performed by taking American accent and British accent as source domain and target domain respectively and vice-versa.
	
	For TIMIT speech corpus male and female speakers are separated into source domain and target domain datasets. TIMIT is a read speech corpus in which a speaker reads a prompt in front of the microphone. It consists of a total of 6300 sentences, 10 sentences spoken by each of the	630 speakers for 8 major dialect regions of the United States of America. It consists
	a total of 3,696 training utterances sampled at 16kHz, excluding all SA utterances because they create
	a bias in the dataset. The training set consists of 438 male speakers and 192 female speakers. The core test set is used to report the results. It consists of 16 male speakers and 8 female speakers from all of the 8 dialect regions. 
	
    For the Voxforge dataset American accent speech and British accent speech are taken as two separate domains. Voxforge is a multi-accent speech dataset with 5 sec speech samples sampled at 16 KHz. Speech samples are recorded by users with their own microphones therefore quality varies significantly among samples. Voxforge corpus has 64 hrs of American accent speech and 13.5 hrs of British accent speech totaling to 83 hrs of speech. Results are reported on 400 utterances each for both the accents. Alignments are obtained by using HMM-GMM acoustic model trained using Kaldi~\cite{povey2011kaldi}.
    \vspace{-0.15in}
	\subsection{Input Features}
	\label{sec:features}
	\vspace{-0.08in}
	
	Raw speech features are obtained by using a rectangular window of size 10 ms on raw speech with a frame shift of 10 ms. A context of 31 frames is added to windowed speech features to get a total of 310 ms of context dependent raw speech features. These context dependent raw speech features are mean and variance normalized to obtain final features.
	\vspace{-0.15in}
	\subsection{Model Description}
	
	\label{sec:model}
	\vspace{-0.08in}
	Feature extractor is a 2 layer convolutional neural network. The first convolutional layer has filter size of 64 with 256 feature maps along with the step size of 31. The second convolutional layer has
	filter size of 15 with 128 feature maps and step size of 1. After each convolutional layer a avg-pool layer is used with pooling size of 2 and a ReLU activation unit. Both the
	senone classifier and domain classifier are 4 layer and 6 layer fully connected neural networks with ReLU activation unit and hidden unit size of 1024 and 2048 for TIMIT and Voxforge respectively. 
	
	The weights are initialized in Glorot ~\cite{glorot2010understanding} fashion. The model is trained with  SGD with momentum ~\cite{sutskever2013importance}. Learning rate is selected during the training using formula $\mu_p=\frac{\mu_0}{(1+\alpha*p)^\beta}$, where $p$ increases linearly from $0$ to $1$ as training progresses, $\mu_0=0.01$, $\alpha=10$, and $\beta=0.75$ . A momentum of $0.9$ is also used. The adaptation parameter $\lambda$ is initialized at $0$ and is gradually changed to $1$ according to the formula $\lambda_p=\frac{2}{1 + \exp(-\gamma*p)}-1$, where $\gamma$ is set to $10$ as suggested in ~\cite{dann}. Domain labels are switched $10\%$ of the times to stabilize the adversarial training.  
    \vspace{-0.15in}
\section{RESULTS}
\vspace{-0.08in}
This section presents the results of experiments performed to evaluate the impact of adversarial training for unsupervised domain adaptation. The experiments specifically study the acoustic variabilities like speaker gender and accent using TIMIT and Voxforge speech corpus respectively. Sec. \ref{sec:male-female} presents the domain adaptation experiments for speaker gender variability. Due to insufficient labeled female speech data in TIMIT corpus domain adaptation, experiments are performed only for male speech as the source domain and female speech as target domain. Sec. \ref{sec:american-british} discusses the experiments performed to mitigate speaker accent variability using American and British accent speech data in Voxforge. Experiments are performed by taking American accent as source domain and British accent as target domain and vice versa. Additional experiments are also performed by training the acoustic model on the labeled data from both the domains which work as the lower limit for the achievable WER. In Table \ref{table:1} and \ref{table:2}, DANN represents the domain adapted acoustic model using labeled data from the source domain and unlabeled data from the target domain and NN represents the acoustic model trained on the labeled data from the source domain only. We also trained a fully connected DNN using MFCC features, with two layers for the feature extractor and three each for the senone and domain classifiers. Each fully connected layer comprised of 1024 nodes and the total number of parameters is similar to that of the model described in Sec. \ref{sec:model}.  
\vspace{-0.15in}
\subsection{Domain adaptation for male and female speech domains in TIMIT corpus}
\label{sec:male-female}
\vspace{-0.08in}
The first two rows in Table \ref{table:1} list the PER results for the acoustic model trained on labeled data from both the domains with no domain adaptation. This acoustic model gives the best results and is the lower limit for the PER. Rows 3 and 4 give the acoustic model trained on labeled data from the male speakers and adapted using unlabeled data from female speakers. Row 3 indicates the effect of domain adaptation on the performance on data from source domain which is male speech in this case. Row number 4 gives the PER for the unadapted and adapted acoustic models for data from target domain which is female speech in this case. With male and female speech as source and target domains respectively, and using MFCC features as input, the PER dropped from 33.825\% to 31.375\% upon applying domain adaptation, indicating a higher absolute accuracy, but a poorer relative improvement compared to raw speech. 
\vspace{-0.15in}
\subsection{Domain adaptation for American and British accents domains in Voxforge corpus}
\label{sec:american-british}
\vspace{-0.08in}
Rows 1 and 2 in Table \ref{table:2} are the WER values for the acoustic model trained on labeled data from both the domains and without any domain adaptation. These values correspond to lower limits for the WER for both the domains. Row number 3 and 4 represents the effect of domain adaptation on the performance of acoustic model on the data from source domain which is American and British respectively. Rows number 5 and 6 gives the WER for target domain data on unadapted and adapted acoustic models. With American and British speech as source and target domains respectively, and using MFCC features, the WER dropped from 24.19\% to 23.73\% upon applying domain adaptation. These results indicate a poorer absolute accuracy as well as relative improvement compared to the raw speech experiments reported in Table \ref{table:2}.   	
		
		
		
			
			
			
			
			
			
			

			
			
		
		
		
	
	\begin{table}[t]
		\fontsize{7}{10}\selectfont
		\centering
		
		\centering
		
		\label{tab:per}
		
		\begin{tabular}{||c c c c c||} 
			
			\hline
			
			Labeled source data & Unlabeled target data & Test data  & NN &DANN \\ [0.5ex] 
			
			\hline\hline
			
                       Male + Female& & Male & 21.25 & \\
                       Male + Female& & Female & 23.21 & \\
           Male & Female & Male  & 24.63& 25.37\\
            
            Male & Female &Female & 37.20& \textbf{32.26} \\ 
			
			
			
			
			\hline
			
		\end{tabular}
		
		\caption{\% PER for for acoustic model trained on supervised data from source domain and unsupervised data from target domain for TIMIT corpus taking male speech as source and female speech as target.}

        \label{table:1}
	\end{table}
	
	\begin{table}[t]
		\fontsize{7}{10}\selectfont
		\centering
		
		\label{tab:wer}
		
		\begin{tabular}{||c c c c c||} 
			
			\hline
			
			Labeled source data & Unlabeled target data & Test data  & NN & DANN \\ [0.5ex] 
			
			\hline\hline
			
            American + British& & American & 10.87 & \\
            American + British& & British & 15.01& \\
            American& British &American   & 11.50 & 16.53 \\
			British& American &British  & 18.41 & 19.62 \\
             American.& British &British  & 28.11 & \textbf{23.10} \\
			
			British& American &American & 23.37 & \textbf{23.16} \\
            

			
			
			\hline
			
		\end{tabular}
		
		\caption{\% WER for acoustic models trained on supervised data from source domain and unsupervised data from target domain for Voxforge dataset taking American and British accents as two different acoustic domains. }
		
		\label{table:2}
	\end{table}

		
		
			
			
			
			
			
            

			
			
			
		
		

\vspace{-0.15in}    
    \section{discussion}
    \vspace{-0.08in}
    
    As can be seen from rows 3 and 4 of Table \ref{table:1}, the acoustic variability due to speaker gender results in a 12.57 \% absolute increase in PER when the acoustic model \added[id=Aditay]{(NN)} trained on male speech (source domain) is tested on female speech (target domain). The effect of applying DANN improves the absolute PER performance on the target domain by nearly 5\%. 
    Similarly, domain shifts due to speaker accent cause the target domain performance deteriorate significantly. Table \ref{table:2} reports an absolute degradation of 16.61\% when the acoustic model is trained on American speech (source domain) and tested on British speech (target domain). Upon applying DANN using the unlabeled British speech, the absolute WER drops by nearly 5\%. General trends that appear from our experimental analysis point out that models undergoing adversarial training based unsupervised domain adaptation, improve in performance on the target domain data as opposed to their unadapted counterparts. This improvement however, comes at a cost of drop in source domain accuracy. This observation is not unexpected as the feature extractor module perhaps learns to ignore some domain-specific features in the pursuit of learning invariant representations.

\vspace{-0.15in} 
	\section{conclusion}
    \vspace{-0.08in}
    The paper proposes unsupervised domain invariant features learning directly from raw speech using domain adversarial neural networks. The senone classification model is modified for domain adaptation by using an additional domain classifier and modifying the loss such that the network learns features from raw speech that are sufficiently discriminative for the senone classifier and invariant enough to fool the domain classifier. This experimental study also shows that there is a significant acoustic variability present in the speech signal due to speaker gender and accent which adversely affects the performance of models trained in a domain agnostic setting. The performance loss due to this variability can be alleviated by adversarial training based domain adaptation using unlabeled target domain data. Moreover, the analysis suggests that raw speech features along with a convolutional neural network based feature extractor may be more amenable to an advesarial approach to domain adaptation as opposed to hand-crafted features like MFCC, particularly when large amounts of data is available. The evaluation of the proposed approach is done using two benchmarking datasets: TIMIT for gender based domain shift and Voxforge corpus for using American and British accents for the domain shift. 
	
\section{Acknowledgment}
While the first author was at IIIT-Delhi, this work was partially supported by the Infosys Center for Artificial Intelligence, IIIT-Delhi.
	
    \vspace{-0.15in}
	\bibliographystyle{IEEEbib}
	\bibliography{refs}
	
\end{document}